\documentclass[prd,preprint,tightenlines,floatfix,showpacs,preprintnumbers,nofootinbib,eqsecnum]{revtex4-1}

\pdfoutput=1

\usepackage{color,xcolor}
\usepackage{amssymb}
\usepackage{amsmath}
\usepackage{amsfonts}
\usepackage{bm}
\usepackage{comment}
\usepackage{float}
\usepackage[utf8]{inputenc}
\usepackage{comment}
\usepackage{tabularx}
\usepackage{booktabs} 
\usepackage{siunitx}  
\usepackage{array}
\usepackage{multirow}
\usepackage[pdftex,final]{graphicx}

\colorlet{RED}{red}

\begin{document}
 

\title{\textbf{Exclusive photoproduction of vector toponium within the color dipole picture}}

\author{Victor P. Gonçalves$^{1}$}
\email{barros@ufpel.edu.br}

\author{Bruno D. Moreira$^{2}$}
\email{bduartesm@gmail.com}

\author{Haimon Trebien$^{2}$}
\email{haimontrebien@outlook.com}

\affiliation{
\\
{$^1$\sl Institute of Physics and Mathematics,
Federal University of Pelotas (UFPel),
Postal Code 354, 96010-900, Pelotas, RS, Brazil
}\\
{$^2$\sl Departamento de F\'isica, Universidade do Estado
de Santa Catarina, 89219-710 Joinville, SC, Brazil\vspace{5mm}
}}

\begin{abstract}\vspace{5mm}
The exclusive photoproduction of the toponium vector meson ($\psi_t$)  is investigated within the color dipole model. The overlap function between the light-cone wave function of the vector toponium state and the photon wave function is estimated using the Boosted Gaussian approach. Considering the dominance of small dipoles, the dipole-proton scattering amplitude is estimated in terms of parton distribution functions, and predictions for the photoproduction cross-section in the energy range of  the proposed Large Hadron electron Collider (LHeC) are presented.  Moreover, the cross-sections and rapidity distributions for the exclusive $\psi_t$ photoproduction  in ultraperipheral proton–proton and proton–nucleus  collisions at the Large Hadron Collider (LHC) and Future Circular Collider (FCC) energies are calculated. We find that exclusive $\psi_t$ photoproduction is, in principle,  experimentally feasible   in ultraperipheral $pp$ collisions at the FCC-hh.
\end{abstract}

\maketitle  

\newpage

\section{Introduction}
\label{introduction}

The photoproduction of heavy quarkonium has been extensively investigated in the charmonium and bottomonium sectors, from theoretical and experimental perspectives. In particular, several studies based on the color dipole formalism have successfully described the exclusive photoproduction of the vector $J/\psi$ and $\Upsilon$ meson states, providing an important framework for investigating the gluonic structure of the proton and QCD dynamics in the small - $x$ regime (See, e.g., Refs. ~\cite{Klein:1999qj,Goncalves:2001vs,Frankfurt:2001db,Goncalves:2005yr,Goncalves:2005ge, Lappi:2013am,  Goncalves:2016sqy, Goncalves:2017wgg,  Henkels:2020qvo, Henkels:2020kju, Cepila:2025exl}). These successful applications of the dipole approach naturally motivate its extension to the heaviest quarkonium system - the vector  bound $t\bar{t}$ state - commonly referred to as toponium. This is one of the goals of this study.

The formation of a toponium state in $e^+e^-$ and $pp$ collisions has been proposed 
many years ago \cite{Fadin:1990wx,Fabiano:1993vx}, but only recently observed at the LHC \cite{ATLAS:2019hau,CMS:2025kzt}, with data being consistent with the formation of a pseudoscalar $t\bar{t}$ state ($\eta_t$ state) \footnote{ For recent phenomenological studies about this state, see e.g. Refs.~\cite{ Fuks:2024yjj, Garzelli:2024uhe, Francener:2025tor, Najjar:2025bby, Jiang:2024fyw}.}. Such a result has motivated the studies performed,  e.g., in Refs.~\cite{Bai:2025buy,Goncalves:2025hyx}, which have investigated if the observation of the $\psi_t$ state is also feasible in hadronic collisions.
In particular, Ref.~\cite{Goncalves:2025hyx} has pioneered the calculation of the cross-section for the exclusive $\psi_t$ production in photon-hadron interactions and provided, for the first time, predictions for the formation of this state in ultraperipheral hadronic collisions.
Another goal of this study is to perform a similar analysis, but based on a different approach to describe the interaction, which will allow us to estimate the current theoretical uncertainty on the predictions.

As emphasized above, a convenient and powerful framework to describe the exclusive vector meson photoproduction is the color dipole formalism \cite{Nikolaev:1990ja,Nikolaev:1991et,Nemchik:1996cw,Nemchik:1996pp}.
 In this picture, the scattering amplitude can be represented as presented in Fig.~\ref{fig:phit-photoproduction0}, and is interpreted as follows: the incoming photon fluctuates into a quark-antiquark color dipole that scatters off the hadronic target before recombining into the final vector meson state. The two main ingredients of the scattering amplitude are the dipole-target scattering amplitude, which describes the diffractive interaction, and the overlap between the photon and vector meson wave functions, which accounts for the photon fluctuation into the $q\bar{q}$ pair and the subsequent bound-state formation. In this paper, we will extend this approach for the vector toponium production. First, we will estimate the overlap between the light-cone $\psi_t$ wave function and the photon wave function using the  Boosted Gaussian model~\cite{Kowalski:2006hc}. As we will demonstrate in the next section, such a function is dominated by dipoles of very small size $r$. As a consequence,  the description of the dipole-proton scattering amplitude can be performed using the color transparency approximation \cite{Barone:2002cv}, which allows expressing this quantity in terms  of the gluon parton distribution function (PDF). In our analysis we will consider distinct gluon PDFs and provide predictions for the energy dependence of the cross-section for the exclusive $\psi_t$ photoproduction. In particular, we will present results for the kinematical range that is expected to be probed in the proposed Large Hadron electron Collider (LHeC) \cite{LHeCStudyGroup:2012zhm,LHeC:2020van}. 

Exclusive interactions in hadronic colliders also offer a particularly clean environment for studying the photoproduction of heavy vector mesons \cite{Klein:1999qj,Goncalves:2001vs,Frankfurt:2001db}. These processes are experimentally accessed in ultraperipheral collisions (UPCs)~\cite{Baur:2001jj, Bertulani:2005ru, Baltz:2007kq, Contreras:2015dqa}, where the impact parameter exceeds the sum of the radii of the colliding particles. Under these conditions, strong hadronic interactions are effectively suppressed, and the process proceeds predominantly through photon exchange, providing a suitable scenario to investigate photon-hadron (and photon-photon) interactions. The hadron-hadron cross section for vector meson production is evaluated within the Equivalent Photon Approximation (EPA)~\cite{Budnev:1975poe}, in which one of the colliding hadrons acts as a source of quasi-real photons while the other serves as the target. Consequently, the total cross section is factorized into the equivalent photon flux, which describes the intensity of the photon field emitted by the source, and the photon-hadron cross section, where all the underlying QCD dynamics reside. The study of the exclusive vector meson photoproduction in UPCs using the color dipole approach was performed for the first time in Ref.~\cite{Goncalves:2005yr}, and widely used in the literature over the last decades. Here, we extend this approach for the calculation of the exclusive $\psi_t$ photoproduction in ultraperipheral $pp$ and $pPb$ collisions. Such a process will be characterized by the presence of two rapidity gaps and intact incident hadrons in the final state.  In particular, the separation of the associated events is expected to be possible in $pp$ collisions using the forward proton detectors currently operating at the LHC, such as the ATLAS Forward Proton (AFP) \cite{Adamczyk:2015cjy, Tasevsky:2015xya} detector and the CMS--TOTEM Precision Proton Spectrometer (CT-PPS) \cite{CMS:2014sdw}. These detectors allow the identification of exclusive events by measuring the intact outgoing protons and are primarily sensitive to final states with large invariant masses ($M>200$ GeV). Owing to the large mass of the vector toponium state, exclusive $\psi_t$ photoproduction represents a promising channel for future experimental studies at the High-Luminosity LHC (HL-LHC) as well as at next-generation hadron colliders, such as the Future Circular Collider (FCC), where the higher center-of-mass energies are expected to further enhance the production rates.




This paper is organized as follows. In Sec.~\ref{formalism}, we present a brief review of the vector meson production in the color dipole formalism. In particular, we will estimate, for the first time, the overlap between light cone $\psi_t$ wave function  and photon wave function. Next, in Sec.~\ref{results}, we will present our predictions for the energy dependence of the photon-proton cross-section in the energy range of the proposed LHeC. Moreover, we will present our predictions for the total cross-sections and rapidity distributions associated with the exclusive $\psi_t$ photoproduction in $pp$ and $pPb$ collisions at the LHC and FCC energies.   Finally, in Sec.~\ref{sec:summary}, we summarize our main conclusions.

\section{Vector meson production in the color dipole formalism}
\label{formalism}

\begin{figure}[t]
\begin{minipage}{0.6\textwidth} \centerline{\includegraphics[width=1.0\textwidth]{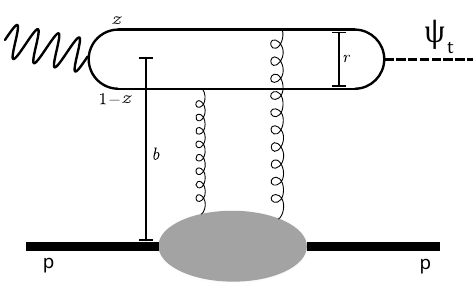}} \end{minipage} \hfill \caption{Scattering amplitude for the exclusive vector toponium photoproduction within the dipole formalism.} 
\label{fig:phit-photoproduction0} 
\end{figure}

In this section we will   discuss in more detail the ingredients needed to calculate the cross-section for the exclusive $\psi_t$ photoproduction. For the proton target case, the  differential cross-section for the exclusive photoproduction of a vector meson $V$, described by the reaction $\gamma p \rightarrow V p$, is given by
\begin{equation}
    \frac{d \sigma^{\gamma p \rightarrow V p}}{d t} =
    \frac{1}{16 \pi }
    \left|\mathcal{A}^{\gamma p \rightarrow V p}(x,\Delta_T)\right|^2,
    \label{amplitude_diff}
\end{equation}
where $t=-\Delta_T^2$ is the squared momentum transfer, $\Delta_T \equiv |\mathbf{\Delta}|$ is the transverse momentum transferred to the proton, and $\mathcal{A}^{\gamma p \rightarrow V p}$ is the scattering amplitude. Within the dipole picture, the incoming quasi-real photon fluctuates into a $q\bar q$ color dipole of transverse size $\mathbf r$ and longitudinal momentum fraction $z$, which subsequently interacts elastically with the target through gluon exchange before recombining into the observed vector meson. The production amplitude, represented in Fig.\ref{fig:phit-photoproduction0}, reads \cite{Kowalski:2003hm}
\begin{equation}
    \mathcal{A}^{\gamma p \rightarrow V p}(x,\Delta_T)
    =
    2i
    \int d^2\mathbf r
    \int_0^1 dz
    \int d^2\mathbf b\,
    (\Psi_V^*\Psi_\gamma)
    e^{-i[\mathbf b-(1-2z)\mathbf r/2]\cdot\mathbf\Delta}
    N(x,\mathbf r,\mathbf b),
    \label{hatta correction}
\end{equation}
where $\mathbf b$ denotes  the impact parameter of the dipole relative to the target center and $N(x,\mathbf r,\mathbf b)$ is the imaginary part of the forward dipole-target scattering amplitude, which describes the interaction. The term $\psi_{V}^{*}\psi_{\gamma}$ is the overlap between the photon and the vector meson wave functions and is represented by the upper part of the diagram. The exponential factor accounts for the non-forward kinematics and ensures the Fourier transformation from impact-parameter space to momentum-transfer space.

The overlap between the transverse photon and vector meson light-front wave functions can be expressed as follows \cite{Forshaw:2003ki, Kowalski:2006hc}
\begin{equation}
(\Psi_V^*\Psi_\gamma)(r,z)
=
e_fe
\frac{N_c}{\pi z(1-z)}
\left[
m_f^2K_0(\epsilon r)\phi(r,z)
-
\left(z^2+(1-z)^2\right)
\epsilon K_1(\epsilon r)
\partial_r\phi(r,z)
\right],
\end{equation}
where $K_0$ and $K_1$ are modified Bessel functions of the second kind,
\begin{equation}
\epsilon^2=z(1-z)Q^2+m_f^2,
\end{equation}
with $Q^2=0$ for photoproduction. Here $m_f$ denotes the top quark mass, $e_f$ is the fractional electric charge of the quark, and $N_c=3$ is the number of colors.
In the Boosted Gaussian (BG) approach, the scalar part of the  wave function $\phi$ is given by~\cite{Kowalski:2006hc,Forshaw:2003ki}
\begin{equation}
\phi(r,z)
=
\mathcal N_T
z(1-z)
\exp\left[
-\frac{m_f^2\mathcal R^2}{8z(1-z)}
-\frac{2z(1-z)r^2}{\mathcal R^2}
+\frac{m_f^2\mathcal R^2}{2}
\right],
\end{equation}
where $\mathcal R$ determines the transverse size of the meson and $\mathcal N_T$ is fixed by the wave-function normalization together with the leptonic decay width. To extend this approach for the $\psi_t$ case, we will follow the procedure outlined in Ref.~\cite{Kowalski:2006hc}. A key ingredient in determining these parameters is the electronic decay width of the vector meson. Following Ref.~\cite{Eichten:1995ch}, it is given by
\begin{equation}
\Gamma(\psi_t \rightarrow e^+ e^-) = \frac{4 N_c \alpha^2 e_q^2}{3} \frac{|R_S(0)|^2}{m_{\psi_t}^2} \left( 1 - \frac{16 \alpha_s}{3\pi} \right).
\label{decay_wd}
\end{equation}
In Eq.~(\ref{decay_wd}), $R_{S}(0)$ denotes the radial wave function at the origin evaluated in the non-relativistic limit. In our analysis, such a quantity will be taken from Ref.~\cite{Bai:2025buy}. Under these assumptions, we obtain $\Gamma(\psi_t \to e^+e^-) = 9.962\text{ keV}$. The resulting parameters for the description of the overlap function between the vector toponium and photon wave functions are presented in Table~\ref{tab:table1}. 

\begin{table}[t] 
\centering 
\begin{tabular}{|c|c|c|c|c|c|} \hline 
\toprule  \quad State  \quad &\quad  $m_V$ (GeV) \quad & \quad  $m_f$ (GeV) \quad & \quad \quad $e_f$ \quad \quad & \quad \quad $N_T$ \quad \quad &  \quad $\mathcal{R}^2$ (GeV$^{-2}$)  \quad\\
\hline
\midrule
$\psi_t$ & 342.1 & 172.4 & $2/3$ & 0.276 & $3.26 \times10^{-3}$ \\ 
\bottomrule 
\hline
\end{tabular} 
\caption{Parameters for the Boosted Gaussian wave function associated with the $\psi_t$ state.} 
\label{tab:table1} 
\end{table}

\begin{figure}[t]
\begin{minipage}{0.6\textwidth} \centerline{\includegraphics[width=1.0\textwidth]{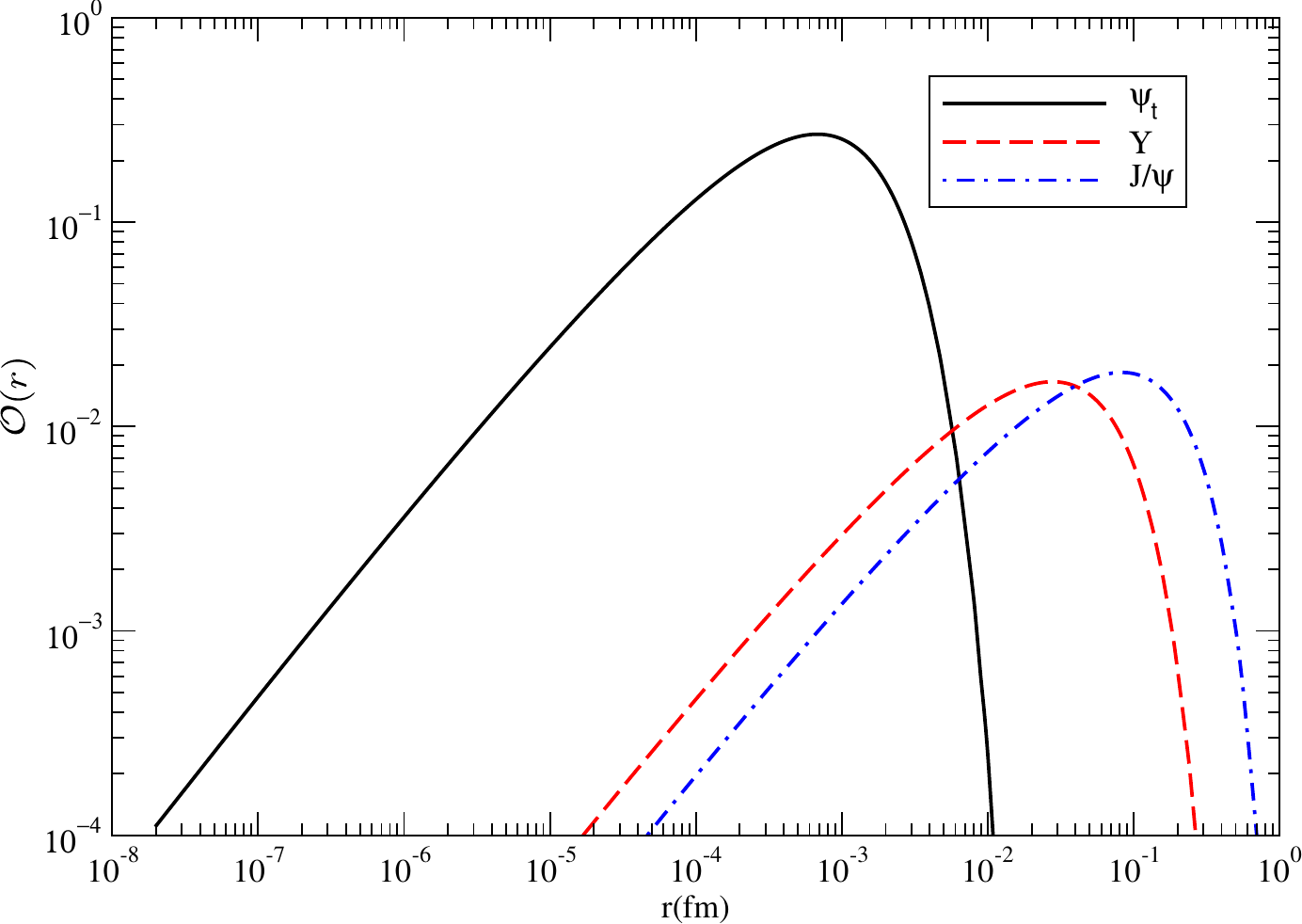}} \end{minipage} \hfill \caption{Overlap between the photon and $\psi_t$ wave functions integrated over z. For comparison, the results for the $J/\psi$ and $\Upsilon$ cases are also shown.} \label{fig:phit-photoproduction1} 
\end{figure}

In Fig.~\ref{fig:phit-photoproduction1}  we present our results for the dependence on the dipole size of the function $\mathcal O(r)$, defined by  
\begin{equation}
\mathcal O(r)
= 2\pi r
\int_0^1dz\,(\Psi_V^*\Psi_\gamma)(r,z),
\end{equation}
which is the overlap function  integrated over the longitudinal momentum fraction $z$, and that
provides direct information on the typical dipole sizes contributing to the scattering amplitude given by Eq.~(\ref{hatta correction}). For completeness, in addition to the result for the overlap between the vector toponium and photon wave functions, we also present the results for the $J/\psi$ and $\Upsilon$ cases. Our results indicate that the distribution associated with the $\psi_t$ meson peaks for a dipole size that is two orders of magnitude smaller than for lighter quarkonia. As a consequence, the scattering amplitude for the exclusive $\psi_t$ photoproduction is predicted to be dominated by very small values of $r$.

The previous result has a direct implication on the description of the dipole-proton scattering amplitude $N(x,\mathbf r,\mathbf b)$.  Such a quantity encodes the information about the hadronic scattering as well as the nonlinear and quantum effects in the proton wave function. Over the recent years, several groups have proposed different approaches to describe $N(x,\mathbf r,\mathbf b)$, based on distinct assumptions for the description of the QCD dynamics at high energies (See, e.g., Ref.~\cite{Morreale:2021pnn}). In general, these approaches differ on the predictions of $N$ for large values of $r$ and small values of $x$. On the other hand, for $r \rightarrow 0$, such a quantity is expected to be proportional to $r^2$,  with this property being called ``color transparency" \cite{Barone:2002cv}, according to which sufficiently small color dipoles interact only weakly with the target. In this regime, 
the dipole amplitude can be expressed by \cite{Kowalski:2003hm}
\begin{equation}
    N(x,\mathbf r,\mathbf b)
    =
    \frac{\pi^2}{2N_c}
    r^2
    \alpha_s(\mu^2)
    xg(x,\mu^2)
    T(b),
    \label{N_amplitude}
\end{equation}
where
\begin{equation}
    \mu^2=\mu_0^2+\frac{4}{r^2}
    \label{mu_scale}
\end{equation}
defines the factorization scale of the gluon distribution \cite{Kowalski:2006hc}. The gluon density $xg(x,\mu^2)$ contains information on the target structure and determines the energy dependence of the dipole-proton scattering amplitude in the color transparency regime. Moreover, the proton impact-parameter profile is assumed to have a Gaussian form,
\begin{equation}
    T(b)=
    \frac{1}{2\pi B_p}
    \exp\left(-\frac{b^2}{2B_p}\right),
    \label{profile_function}
\end{equation}
with $B_p=4~{\rm GeV}^{-2}$ \cite{Kowalski:2006hc},
which satisfies the normalization condition $\int d^2\mathbf b\,T(b)=1$.
This parametrization reproduces the average transverse distribution of gluons inside the proton and has been widely employed in phenomenological analyses of exclusive vector meson production. 

\begin{figure}[t]
\centering
\begin{minipage}{0.495\textwidth}
    \centering
    \includegraphics[width=\textwidth]{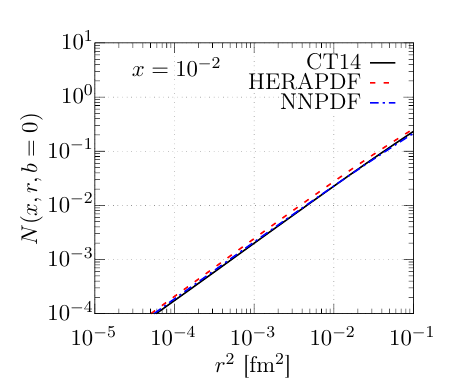}
\end{minipage} \hfill
\begin{minipage}{0.495\textwidth}
    \centering
    \includegraphics[width=\textwidth]{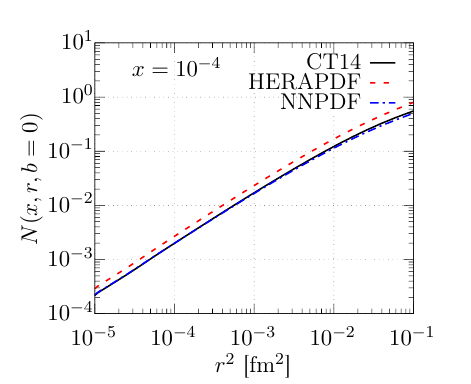}
\end{minipage}
\caption{Dipole-proton scattering amplitude as a function of the dipole size squared $r^2$, evaluated at $b=0$ and different values of $x$. Results derived in the color transparency approximation, Eq.~\eqref{N_amplitude}, assuming different parameterizations for the gluon distribution.}
\label{fig:phit-n_amplitudes}
\end{figure}

In Fig.~\ref{fig:phit-n_amplitudes} we present our predictions for the dipole-proton scattering amplitude as a function of $r^2$, derived assuming $b = 0$, distinct values of $x$ and different parameterizations for the gluon distribution. In particular, in the numerical calculations presented in this work, the gluon density is taken from the leading-order NNPDF3.0~\cite{NNPDF:2014otw}, HERAPDF2.0~\cite{H1:2015ubc}, and CT14~\cite{Dulat:2015mca} parameterizations. Our results indicate that the normalization of $N$ for a given dipole size is larger when $x$ decreases. Moreover, the results are dependent on the parameterization assumed in the calculation, with the difference between the predictions increasing for smaller values of $x$.


\section{Results}
\label{results}
In what follows, we present our predictions for exclusive vector toponium photoproduction in $ep$ collisions and ultraperipheral $pp$ and $p\mathrm{Pb}$ collisions, derived using the color dipole formalism and the assumptions discussed in the previous section. Following Ref.~\cite{Bai:2025buy}, we adopt $m_t = 172.4 \,\mathrm{GeV}$ and $m_{\psi_t} = 342.1 \,\mathrm{GeV}$.

\begin{figure}[t]
\begin{minipage}{0.68\textwidth}
 \centerline{\includegraphics[width=1.0\textwidth]{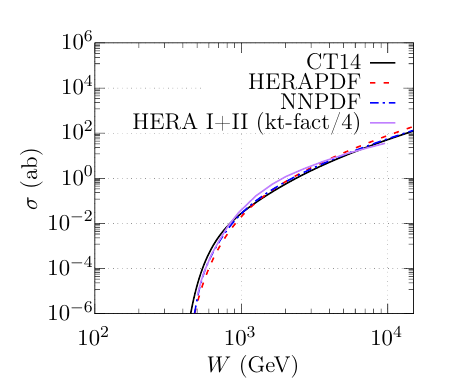}}
\end{minipage} \hfill
\caption{Color dipole predictions for the total cross-section for $\psi_t$ photoproduction as a function of the $\gamma p$ center-of-mass energy $W$, derived considering different PDF sets. For comparison, the prediction derived using the $k_t$-factorization formalism in Ref.~\cite{Goncalves:2025hyx} (HERA I+II), rescaled by a factor of 4 (see text),  is also presented.}
\label{fig:phit-photoproduction2}
\end{figure}

Initially, in  Fig.~\ref{fig:phit-photoproduction2}, we show our predictions for the total $\gamma p \rightarrow \psi_{t} p$ cross-section as a function of the photon-proton center-of-mass energy $W$. The results derived using the color dipole formalism were obtained using the central sets of the leading order CT14, HERAPDF and NNPDF parameterizations for the gluon distribution. It is important to emphasize that we also have considered the replicas provided by the distinct PDF groups and verified that the impact on the predictions is negligible, which is expected since the uncertainty in the gluon distribution at large $Q^2$ is small. As expected from the behavior of the gluon distribution at small $x$ and large values of the hard scale ($Q^2 \approx m^2_{\psi_t}$), we predict a steep increase of the cross-section with the center-of-mass energy. One can observe that the predictions derived considering the distinct gluon PDFs are similar across the entire energy range, spanning from a few to hundreds of $\mathrm{ab}$ in the region considered. In particular, we predict that the cross-section is of the order of $\approx 0.1$ ab in the expected energy range of the LHeC ($W \approx 1000$ GeV).
These color dipole predictions for 
$\sigma(\gamma p \rightarrow \psi_{t} p)$ can be compared with those derived in Ref.~\cite{Goncalves:2025hyx} using the $k_t$-factorization framework. In order to perform this comparison, it is important to consider that in Ref.~\cite{Goncalves:2025hyx} the slope parameter $B_p$, which is not constrained by the formalism used in that reference, was assumed as being equal to $1\,\mathrm{GeV}^{-2}$.
Here, we are assuming
$B_p = 4\,\mathrm{GeV}^{-2}$, as obtained in Ref.~\cite{Kowalski:2006hc} by fitting the HERA data for the $J/\psi$ photoproduction. As this quantity has a direct impact on the normalization of the predictions, to compare the results, we will divide the predictions obtained in Ref.~\cite{Goncalves:2025hyx} by a factor of 4. The associated prediction is denoted by HERA I+II ($k_t$-fact/4) in Fig.~\ref{fig:phit-photoproduction2}. One has that if this rescaling is performed, the color dipole predictions are similar to those derived in Ref.~\cite{Goncalves:2025hyx}.

\begin{figure}[t]
\centering
\begin{minipage}{0.495\textwidth}
    \centering
    \includegraphics[width=\textwidth]{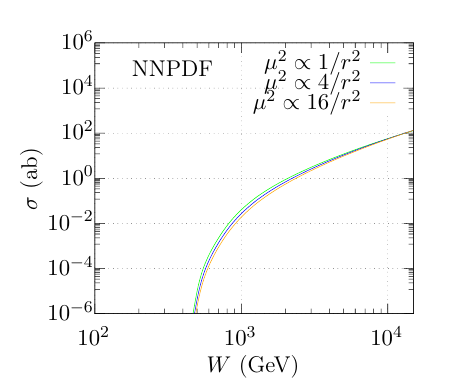}
\end{minipage} \hfill
\begin{minipage}{0.495\textwidth}
    \centering
    \includegraphics[width=\textwidth]{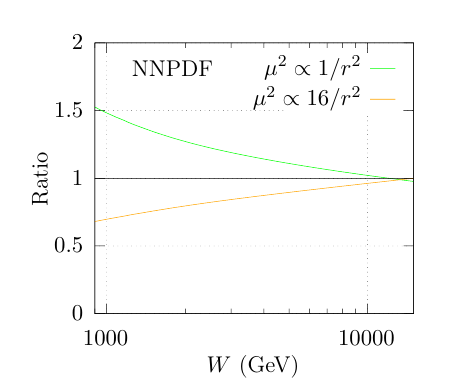}
\end{minipage}
\caption{Cross sections obtained with NNPDF at leading order (LO). Left panel: Total cross section for $\psi_t$ photoproduction as a function of the $\gamma p$ center-of-mass energy $W$ for different factorization scales; Right panel: Ratios of the cross sections for $C = 1$ and $16$ relative to the default case ($C = 4$).}
\label{fig:phit-photoproduction-scale}
\end{figure}

In the above calculations, we have assumed that the hard scale is given by $\mu^{2} = \mu^{2}_{0} + {C}/{r^{2}}$, with $C = 4$, as obtained in Ref.~\cite{Kowalski:2006hc} by fitting the HERA data. In order to estimate the dependence of our predictions on this choice, we have also considered two alternative values for $C$: $C = 1$ and $C = 16$. The associated predictions are presented in Fig.~\ref{fig:phit-photoproduction-scale}, where we compare the predictions obtained with different choices of the factorization scale and  the NNPDF parameterization.  As shown in the left panel the impact of these different choices for $\mu^2$ on the total cross-section is small. To make this comparison clearer, the right panel  presents the ratios of the cross sections for $C = 1$ and $C = 16$ relative to the central case ($C = 4$). The relative differences are of the order of 50\% for $W \approx 1000$ GeV and become noticeably smaller with the increase of the center-of-mass energy.

In what follows we will consider the exclusive $\psi_t$ photoproduction in 
ultraperipheral collisions, where the photon-hadron cross section is probed at larger center-of-mass energies than in $ep$ collisions at the LHeC~\cite{Baltz:2007kq}. In ultraperipheral collisions the hadronic cross-section is obtained by convoluting the photon-hadron cross section with the equivalent-photon flux generated by the fast-moving charge (proton or nucleus in our case). In the equivalent photon approximation, the electromagnetic field of a relativistic charge is treated as a flux of quasi-real photons \cite{Baur:2001jj,Bertulani:2005ru}. As a consequence, the rapidity distribution for the exclusive photoproduction of a vector toponium in hadron-hadron collisions is therefore written as
\begin{equation}
\frac{d\sigma^{h_{1}h_{2}\rightarrow  h_{1} \otimes \psi_t \otimes h_{2}}}{dY}
=
n_{h_{1}}(\omega_+)
\sigma_{\gamma h_{2}\rightarrow \psi_t\otimes h_2}(W_+)
+
n_{h_{2}}(\omega_-)
\sigma_{\gamma_{h_{1}\rightarrow \psi_t\otimes h_1}}(W_-),
\label{eq:UPCrapidity}
\end{equation}
where $Y$ is the rapidity of the vector meson.
The photon energies associated with the two beam directions are
\begin{equation}
\omega_{\pm}
=
\frac{M_V}{2}e^{\pm Y},
\end{equation}
where $W_{\pm}^2 = 2\omega_{\pm}\sqrt{s_{NN}}$, is the photon-hadron center-of-mass energy. 
In our analysis, we will assume that the equivalent-photon flux of a nucleus with charge $Z$ is described by \cite{Baur:2001jj,Bertulani:2005ru}, 
\begin{equation}
n(\omega)
=
\frac{2Z^2\alpha_{\rm em}}{\pi\omega}
\left[
\xi K_0(\xi)K_1(\xi)
-
\frac{\xi^2}{2}
\left(
K_1^2(\xi)-K_0^2(\xi)
\right)
\right],
\label{eq:WWflux}
\end{equation}
where $\xi={\omega b_{\rm min}}/{\gamma_L}$,
$K_0$ and $K_1$ are modified Bessel functions of the second kind, $\gamma_L$ is the Lorentz factor of the beam, and $b_{\rm min}$ is the minimum impact parameter, commonly taken as $b_{\rm min}=R_A + R_p$ for nucleus--proton ultraperipheral collisions to suppress hadronic interactions.  For the photon flux associated with the proton, we will assume that it is given by~\cite{Drees:1988pp}
\begin{equation}
n_p(\omega) = \frac{\alpha_{\text{em}}}{2\pi} \left[ 1 + \left( 1 - \frac{2\omega}{\sqrt{s}} \right)^2 \right] \left( \ln\Omega - \frac{11}{6} + \frac{3}{\Omega} - \frac{3}{2\Omega^2} + \frac{1}{3\Omega^3} \right),
\end{equation}
where $\Omega = 1 + [(0.71\text{ GeV}^2) / Q_{\text{min}}^2]$ and $Q_{\text{min}}^2 = \omega^2 / [\gamma_L^2 (1 - 2\omega/\sqrt{s})] \approx (\omega/\gamma_L)^2$.
Comparing the corresponding expressions, we observe that the nuclear photon flux is enhanced by a factor of $Z^2$ ($82^2$ for lead) relative to the proton flux. Consequently, the cross sections for nuclear interactions are significantly larger than those in proton-proton collisions. Furthermore, due to this $Z^2$ enhancement, in proton-nucleus collisions the hadronic cross-section is mainly determined by photon-proton interactions.

\begin{figure}[t]
\centering
\begin{minipage}{0.495\textwidth}
    \centering
    \includegraphics[width=\textwidth]{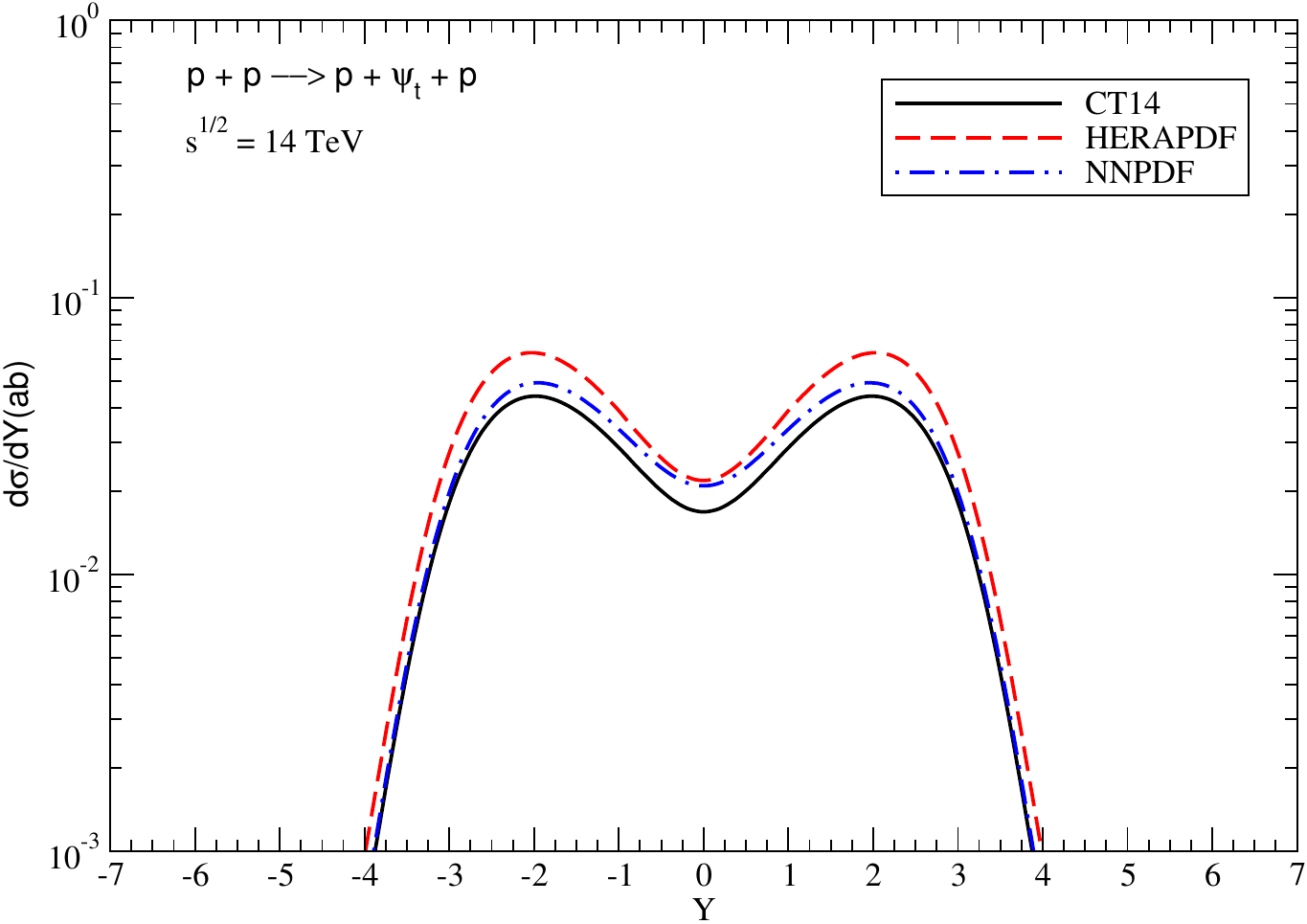}
\end{minipage} \hfill
\begin{minipage}{0.495\textwidth}
    \centering
    \includegraphics[width=\textwidth]{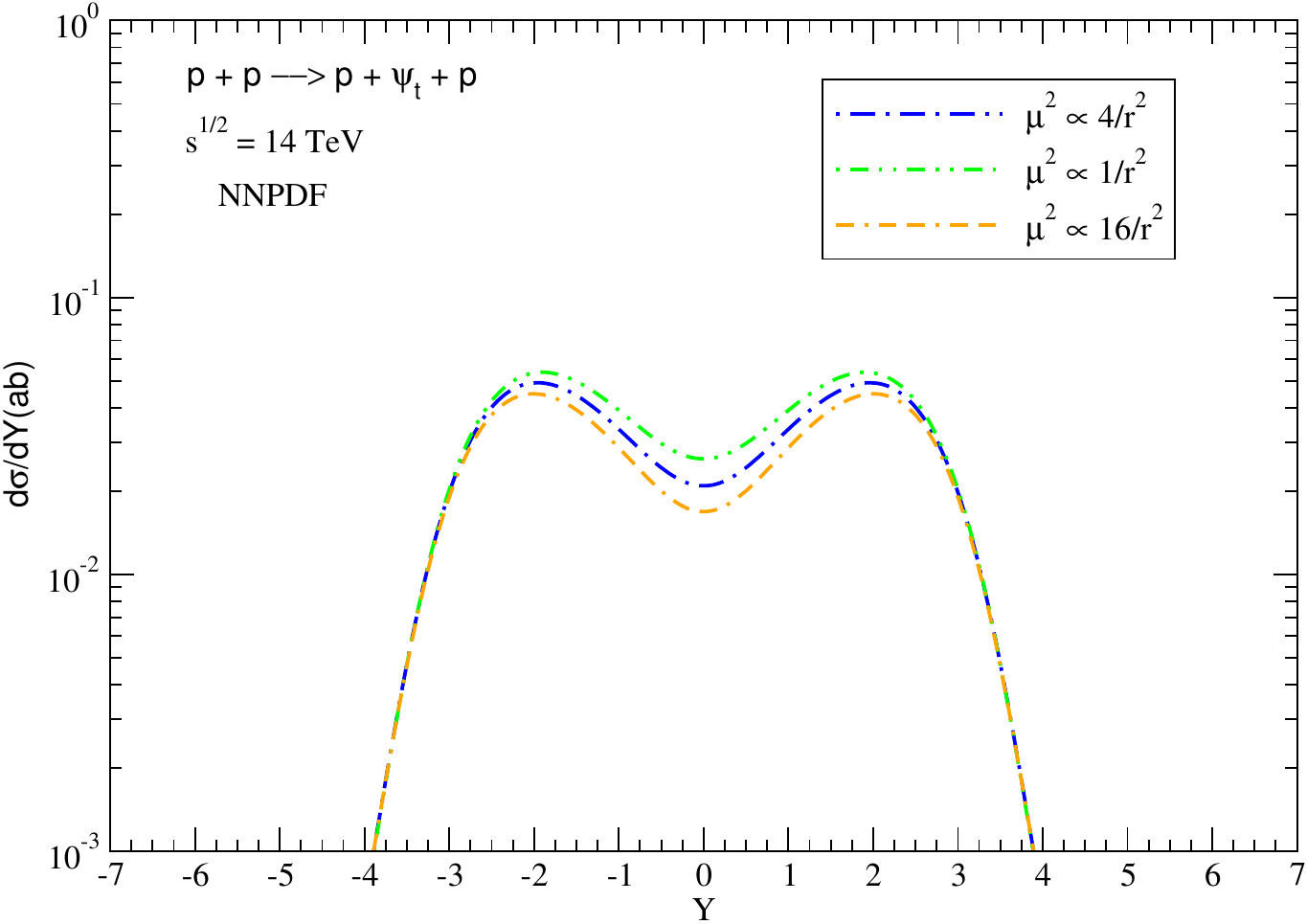}
\end{minipage}
\begin{minipage}{0.495\textwidth}
    \centering
    \includegraphics[width=\textwidth]{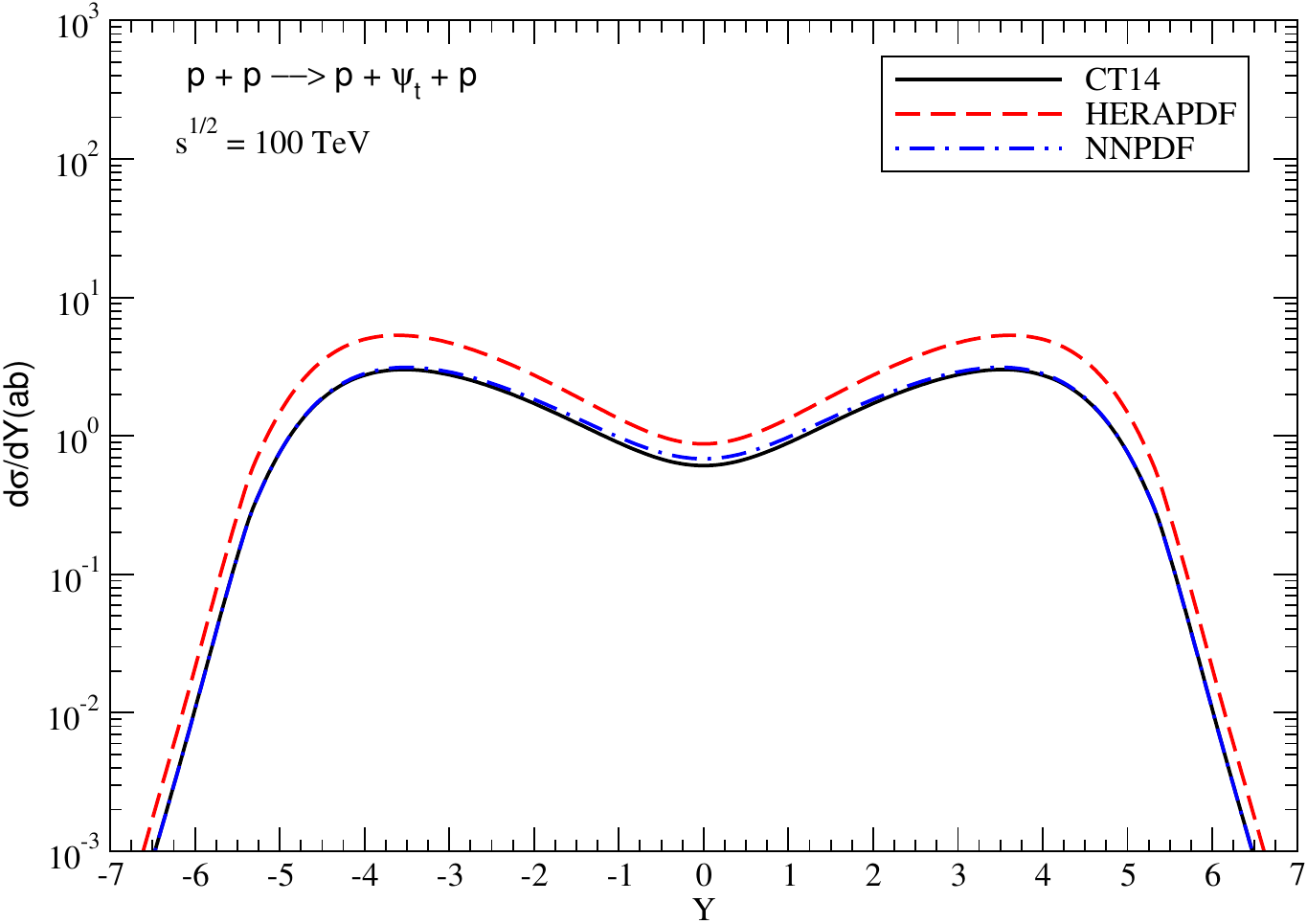}
\end{minipage} \hfill
\begin{minipage}{0.495\textwidth}
    \centering
    \includegraphics[width=\textwidth]{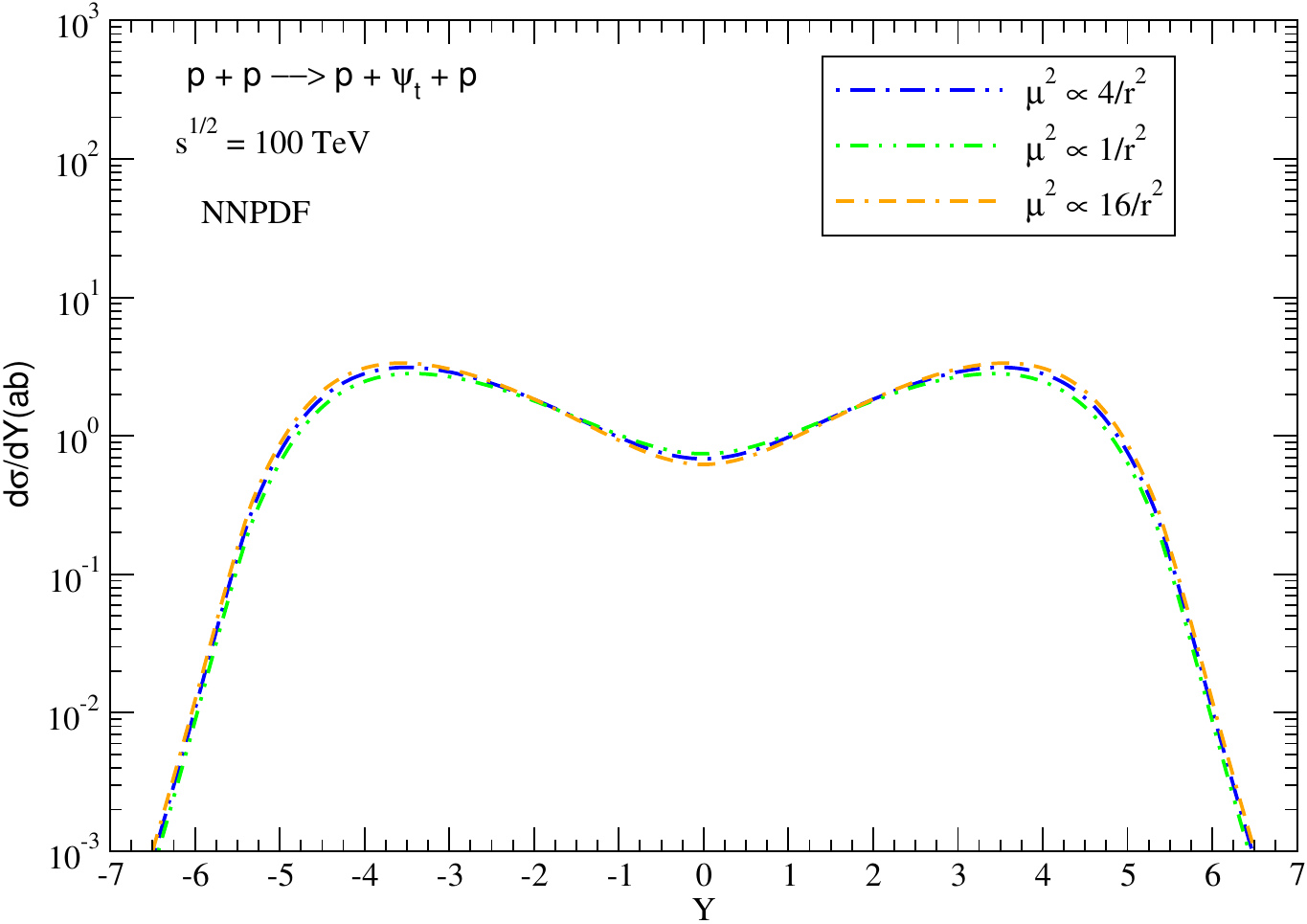}
\end{minipage}
\caption{Rapidity distribution for the exclusive $\psi_t$ photoproduction in $pp$ collisions at the LHC (upper panels) and FCC (lower panels) energies.}
\label{fig:phit-photoproduction4}
\end{figure}

In Fig.~\ref{fig:phit-photoproduction4} we present our results for the rapidity distributions  associated with the  exclusive vector toponium photoproduction in $pp$ collisions at the LHC (upper panels) and FCC (lower panels) energies. In the left panels, we compare the predictions obtained using different  PDF sets. It can be observed that HERAPDF yields the largest normalization, whereas CT14 one gives the smallest normalization in both cases. Furthermore, at LHC energies (upper left panel), the NNPDF prediction is close to the CT14 one at forward and backward rapidities, while around midrapidity ($Y \approx 0$) the NNPDF prediction is similar to the HERAPDF one. On the other hand, at FCC energies (lower left panel), the  NNPDF and CT14 predictions exhibit nearly identical normalizations. As expected, the rapidity distributions are symmetric regarding $Y = 0$ due to the identical photon fluxes of the colliding protons. Additionally, the distributions become broader as the center-of-mass energy ($\sqrt{s}$) increases. In the right panels, we show predictions obtained with the NNPDF set for different choices of the factorization scale $\mu^2$. A small impact is observed around central rapidities at LHC energies, and these differences become even smaller at FCC energies. This behavior is expected from the dependence of the photon-proton cross-section on $\mu^2$, discussed above and presented in Fig.~\ref{fig:phit-photoproduction-scale}.  We have quantified the magnitude of  these scale uncertainties, and estimated that  the main differences are concentrated around $Y = 0$ and LHC energy, reaching up to approximately 20\% relative to the default case ($C =4$).


\begin{figure}[t]
\centering
\begin{minipage}{0.495\textwidth}
    \centering
    \includegraphics[width=\textwidth]{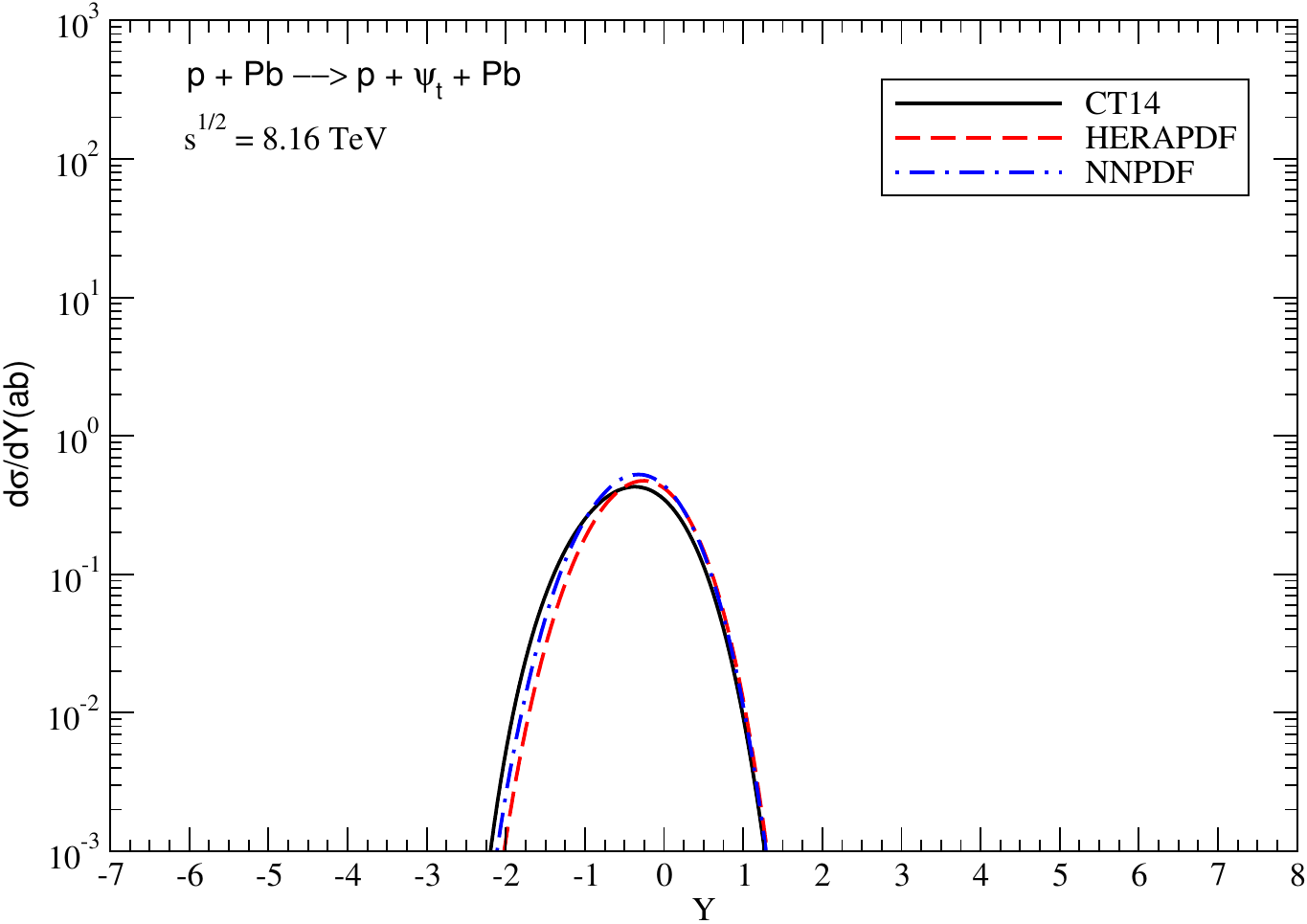}
\end{minipage} \hfill
\begin{minipage}{0.495\textwidth}
    \centering
    \includegraphics[width=\textwidth]{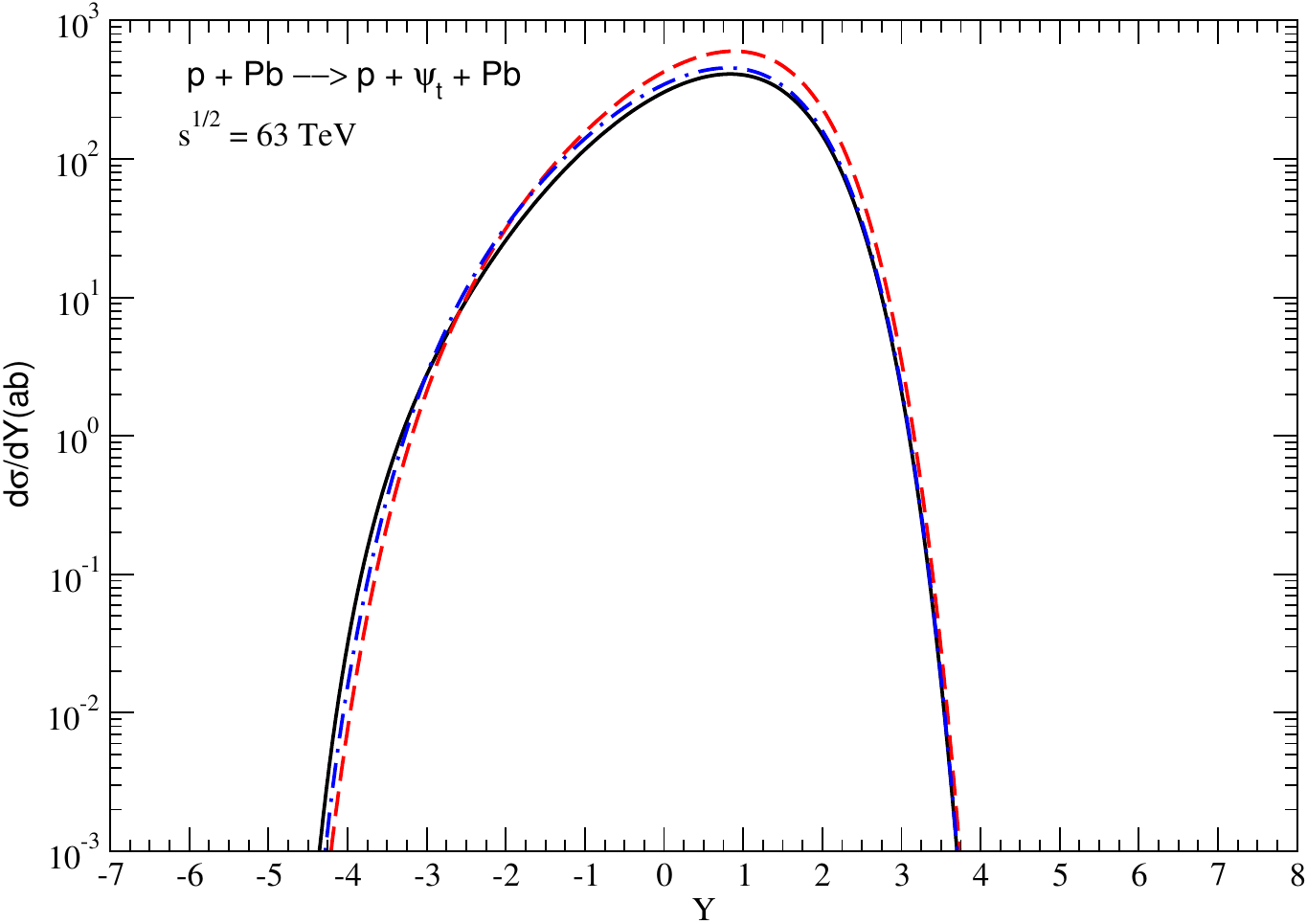}
\end{minipage}
\caption{Rapidity distributions for the exclusive $\psi_t$ photoproduction in $pPb$ collisions at the  LHC (left panel) and FCC (right panel) energies.}
\label{fig:phit-photoproduction7}
\end{figure}

In Fig.~\ref{fig:phit-photoproduction7}, we show our results considering  ultraperipheral $p\mathrm{Pb}$ collisions at the LHC (left panel) and FCC (right panel) energies. As discussed above, as the nuclear photon flux is enhanced by a factor of $Z^2$, the hadronic cross-section is dominated by photon-proton ($\gamma p$) interactions, which implies asymmetric rapidity distributions in $p\mathrm{Pb}$ collisions. One has that as the collision energy increases (from LHC to FCC), both the normalization and the width of the rapidity distributions increase. Furthermore, the main differences among the PDF models occur around the peak and in the backward-rapidity region. As in the $pp$ case, we have also estimated the dependence of our predictions on the choice of the factorization scale and verified that the largest scale uncertainties  are observed at backward rapidities, where smaller photon-proton center-of-mass energies are probed, and are of the order of 60\% for $Y \approx -2$.


\begin{table}[t]
    \centering
    \resizebox{\linewidth}{!}{%
    \begin{tabular}{|ll|cc|cc|}
        \hline
        & & \multicolumn{2}{c|}{$\sqrt{s_{pp}} = 14\text{ TeV}$ ($\mathcal{L}_{\text{int}} = 3\text{ ab}^{-1}$)} & \multicolumn{2}{c|}{$\sqrt{s_{pp}} = 100\text{ TeV}$ ($\mathcal{L}_{\text{int}} = 30\text{ ab}^{-1}$)} \\ 
        \shortstack[l]{\hspace*{3pt}\textbf{Rapidity}\\\hspace*{3pt}\textbf{Range}} & \textbf{PDF} & $\sigma$ (ab) & $N_{\text{events}}$ & $\sigma$ (ab) & $N_{\text{events}}$ \\
        \hline
        \multirow{3}{*}{\hspace*{3pt}\textbf{Full}} & CT14  & 0.1997 & 0.60 & 18.3814 & 551.4 \\
                      & HERAPDF  & 0.2828 & 0.85 & 31.4255 & 942.8 \\
                      & NNPDF  & 0.2270 & 0.68 & 19.2850 & 578.6 \\
        \hline
        \multirow{3}{*}{\shortstack[l]{\hspace*{3pt}\textbf{ALICE}\\[3pt]\hspace*{3pt}($-2 < y < 2$)}} & CT14  & 0.1181 & 0.35 & 3.9339 & 118.0 \\
                                                       & HERAPDF  & 0.1620 & 0.49 & 5.9811 & 179.4 \\
                                                       & NNPDF  & 0.1371 & 0.41 & 4.2982 & 128.9 \\
        \hline
        \multirow{3}{*}{\shortstack[l]{\hspace*{3pt}\textbf{LHCb}\\[3pt]\hspace*{3pt}($2 < y < 4.5$)}} & CT14  & 0.0408 & 0.12 & 6.3511 & 190.5 \\
                                                      & HERAPDF  & 0.0604 & 0.18 & 11.0472 & 331.4 \\
                                                      & NNPDF  & 0.0449 & 0.13 & 6.6154 & 198.5 \\
        \hline
        \hline
        & & \multicolumn{2}{c|}{$\sqrt{s_{p\text{Pb}}} = 8.16\text{ TeV}$ ($\mathcal{L}_{\text{int}} = 1.2\text{ pb}^{-1}$)} & \multicolumn{2}{c|}{$\sqrt{s_{p\text{Pb}}} = 63\text{ TeV}$ ($\mathcal{L}_{\text{int}} = 29\text{ pb}^{-1}$)} \\
        \shortstack[l]{\hspace*{3pt}\textbf{Rapidity}\\\hspace*{3pt}\textbf{Cut}} & \textbf{PDF} & $\sigma$ (ab) & $N_{\text{events}}$ & $\sigma$ (ab) & $N_{\text{events}}$ \\
        \hline
        \multirow{3}{*}{\hspace*{3pt}\textbf{Full}} & CT14  & 0.6090 & $< 10^{-5}$ & 1003.94 & 0.029 \\
                      & HERAPDF  & 0.6119 & $< 10^{-5}$ & 1447.21 & 0.042 \\
                      & NNPDF  & 0.6953 & $< 10^{-5}$ & 1129.14 & 0.033 \\
        \hline
        \multirow{3}{*}{\shortstack[l]{\hspace*{3pt}\textbf{ALICE}\\[3pt]\hspace*{3pt}($-2 < y < 2$)}} & CT14  & 0.6953 & $< 10^{-5}$ & 943.951 & 0.027 \\
                                                       & HERAPDF  & 0.6118 & $< 10^{-5}$ & 1358.36 & 0.039 \\
                                                       & NNPDF  & 0.6950 & $< 10^{-5}$ & 1063.31 & 0.031 \\
        \hline
        \multirow{3}{*}{\shortstack[l]{\hspace*{3pt}\textbf{LHCb}\textsuperscript{*}\\[3pt]\hspace*{3pt}($2 < y < 4.5$)}} & CT14  & 0.0006 & $< 10^{-5}$ & 48.0625 (11.8997) & 0.0014 (0.0003) \\
                                                          & HERAPDF  & 0.0001 & $< 10^{-5}$ & 76.1359 (12.6775) & 0.0022 (0.0004) \\
                                                          & NNPDF  & 0.0003 & $< 10^{-5}$ & 51.5767 (14.2212) & 0.0015 (0.0004) \\
        \hline
        \multicolumn{6}{|l|}{\footnotesize \hspace*{3pt}\textsuperscript{*}For $p\text{Pb}$, values in parentheses denote cross sections and expected events in [$-4.5 < y < -2$].} \\
        \hline
    \end{tabular}%
    }
        \caption{Calculated cross-sections ($\sigma$) and expected number of events ($N_{\text{events}} = \sigma \times \mathcal{L}_{\text{int}}$) for exclusive vector toponium ($\psi_t$) photoproduction in ultraperipheral $pp$ and $p\text{Pb}$ collisions. For $pp$ at $14$~TeV (HL-LHC), we assume $\mathcal{L}_{\text{int}} = 3\text{ ab}^{-1}$ \cite{HL-LHC:2019}; for $pp$ at $100$~TeV (FCC-hh), $\mathcal{L}_{\text{int}} = 30\text{ ab}^{-1}$ \cite{FCC:2019hh}. For $p\text{Pb}$ at $8.16$~TeV (HL-LHC), we adopt the integrated luminosity $\mathcal{L}_{\text{int}} = 1.2\text{ pb}^{-1} = 1.2\times 10^{-6}\text{ ab}^{-1}$~\cite{Citron:2018nin}; for $p\text{Pb}$ at $63$~TeV (FCC-hh), we consider the  luminosity $\mathcal{L}_{\text{int}} = 29\text{ pb}^{-1} = 2.9\times 10^{-5}\text{ ab}^{-1}$~\cite{FCC:2019hh,Dainese:2019FCCAA}. }
\label{total_cross_sections}    
\end{table}

Finally, in Table~\ref{total_cross_sections}, we present the total cross sections and expected number of events ($N_{\text{events}}$) for $pp$ and $p\text{Pb}$ collisions at both colliders, assuming the expected integrated luminosities for the High-Luminosity LHC (HL-LHC) and FCC-hh. Our results are shown for the full rapidity coverage as well as with rapidity cuts corresponding to the ALICE (central) and LHCb (forward) acceptance regions. As can be observed, the expected number of events in $p\text{Pb}$ collisions at both HL-LHC and FCC-hh is negligible ($N_{\text{events}} \ll 1$), owing to the lower integrated luminosities inherent to heavy-ion operations. Consequently, the search for vector toponium photoproduction in $p\text{Pb}$ mode is experimentally unfeasible. Conversely, in  $pp$ collisions at the FCC-hh ($\sqrt{s} = 100\text{ TeV}$), the combination of enhanced cross-sections and high integrated luminosity ($\mathcal{L}_{\text{int}} = 30\text{ ab}^{-1}$) yields several hundreds of events (ranging from $\approx 550$ to $940$ in the full rapidity range), demonstrating that the discovery of vector toponium is, in principle, feasible in $pp$ collisions at this future collider.

\section{Summary}
\label{sec:summary}
The recent observation of a substantial excess of events near the kinematical $t\bar{t}$ threshold  in $pp$ collisions at the LHC, which are consistent with the formation of a pseudoscalar $\eta_t$ toponium state, has motivated a more profound analysis of its properties as well as the investigation of whether other toponium states could also be measured.
In particular, the possibility of the formation of a vector toponium state has been discussed in the literature, with results from different studies indicating that its measurement at the LHC would be a hard task, but feasible at the FCC-hh. Such results have motivated the analysis performed in this paper, where we have extended the application of the color dipole formalism, which is a successful approach used to describe quarkonium photoproduction, to vector toponium photoproduction.

In this work, we investigated exclusive vector toponium ($\psi_t$) photoproduction in $ep$ collisions and ultraperipheral $pp$ and $p\text{Pb}$ collisions within the color dipole picture, using a Boosted Gaussian wave function specifically determined for the $\psi_t$ state, assuming the color transparency approximation and distinct parameterizations for the gluon distribution. The photon-proton cross-section has been estimated, and predictions for the expected LHeC energy were presented. Moreover, we have demonstrated that the color dipole predictions agree with those derived in Ref.~\cite{Goncalves:2025hyx} using a distinct approach, if a same slope is considered in the calculations. Moreover, our results indicate that observing $\psi_t$ photoproduction at the HL-LHC (in both $pp$ and $p\text{Pb}$ modes) and in $p\text{Pb}$ collisions at the FCC  is experimentally unfeasible due to small rates ($N_{\text{events}} \ll 1$). Conversely, $pp$ collisions at the FCC-hh ($\sqrt{s} = 100\text{ TeV}$) offer, in principle, a viable discovery channel, yielding hundreds of events for an integrated luminosity of $30\text{ ab}^{-1}$.


\section*{Acknowledgments}
 VPG was partially financed by the Brazilian funding agencies CNPq, FAPERGS and INCT-FNA  (Process No. 408419/2024-5). B.D.M. and H.T. were partially supported by FAPESC (Process No. 2025TR001585).

\bibliography{bib}

\end{document}